\documentstyle[epsfig]{jaa}

\DeclareMathAlphabet{\mathsc}{OT1}{cmr}{m}{sc}
\def\testbx{bx}%
\DeclareRobustCommand{\ion}[2]{%
\relax\ifmmode
\ifx\testbx\f@series
{\mathbf{#1\,\mathsc{#2}}}\else
{\mathrm{#1\,\mathsc{#2}}}\fi
\else\textup{#1\,{\mdseries\textsc{#2}}}%
\fi}

\def\spose#1{\hbox to 0pt{#1\hss}}
\def\simlt{\mathrel{\spose{\lower 3pt\hbox{$\mathchar"218$}}
     \raise 2.0pt\hbox{$\mathchar"13C$}}}
\def\simgt{\mathrel{\spose{\lower 3pt\hbox{$\mathchar"218$}}
     \raise 2.0pt\hbox{$\mathchar"13E$}}}

\def\aj{AJ}                   
\def\apj{ApJ}                 
\def\aap{A\&A}                
\def\mnras{MNRAS}             

\begin{document}
\title[Northern radio surveys and shocks and magnetic fields in clusters]
{LOFAR and APERTIF surveys of the radio sky: \\
probing shocks and magnetic fields in galaxy clusters}
\author[Huub R\"ottgering]%
{Huub  R\"ottgering$^{1}$\thanks{e-mail:rottgering@strw.leidenuniv.nl},%
Jose Afonso${^2}$,%
Peter   Barthel${^3}$,
\newauthor%
Fabien  Batejat${^4}$,%
Philip  Best${^5}$,%
Annalisa        Bonafede${^6}$,
\newauthor%
Marcus          Br\"uggen${^6}$,%
Gianfranco      Brunetti${^7}$,%
Krzysztof       Chy\.zy${^8}$,
\newauthor%
John            Conway${^9}$,%
Francesco       De Gasperin$^{10}$,%
Chiara          Ferrari$^{11}$,
\newauthor%
Marijke         Haverkorn$^{1,12,17}$,%
George          Heald$^{12}$,%
Matthias        Hoeft$^{13}$,
\newauthor%
Neal            Jackson$^{14}$,%
Matt            Jarvis$^{15}$,%
Louise          Ker$^{5}$,
\newauthor%
Matt            Lehnert$^{16}$,%
Giulia          Macario$^{7}$,%
John            McKean$^{12}$,
\newauthor%
George          Miley${^1}$,%
Raffaella       Morganti$^{3,12}$,%
Tom             Oosterloo$^{3,12}$,
\newauthor%
Emanuela        Orr\`u$^{17}$,%
Roberto         Pizzo$^{12}$,%
David           Rafferty${^1}$,
\newauthor%
Alexander       Shulevski${^3}$,%
Cyril           Tasse$^{16}$,%
Ilse            van Bemmel$^{12}$,
\newauthor%
Bas             van der Tol${^1}$,%
Reinout         van Weeren${^1}$,%
Marc           Verheijen${^3}$,
\newauthor%
Glenn           White$^{18}$,
Michael         Wise$^{12}$, on behalf of the LOFAR collaboration\\
$^{1}$Leiden Observatory, Leiden University, PO Box 9513, 2300 RA Leiden, The Netherlands     \\
$^{2}$Observat\'orio Astron\'omico de Lisboa, Faculdade de Ci\^encias, \\ Universidade de Lisboa, Tapada da Ajuda, 1349-018 Lisbon, Portugal\\
$^{3}$Kapteyn Instituut, Landleven 12, 9747 AD Groningen, The Netherlands        \\
$^{4}$Chalmers University of Technology, Onsala Space Observatory, SE 439 92 Onsala, Sweden \\
$^{5}$Royal Observatory, Blackford Hill, Edinburgh EH9 3HJ, UK     \\
$^{6}$Jacobs University Bremen, Campus Ring 1, 28759 Bremen, Germany       \\
$^{7}$INAF, Istituto di Radioastronomia, Via P Gobetti 101, IT 40129, Bologna, Italy       \\
$^{8}$Jagiellonian University, ul. Orla 171 30-244 Krak\'ow POLAND \\ 
$^{9}$Chalmers University of Technology, Onsala Space Observatory, SE 439 92 Onsala, Sweden \\
$^{10}$Max-Planck-Institut f\"ur Astrophysik, Karl-Schwarzschildstra{\ss}e 1, 85741 Garching, Germany\\
$^{11}$UNS, CNRS UMR 6202 Cassiop\'ee, Observatoire de la Cote d'Azur, Nice, France \\
$^{12}$ASTRON, PO Box 2, 7990 AA Dwingeloo, The Netherlands    \\
$^{13}$Th\"uringer Landessternwarte, Tautenburg      \\
$^{14}$Jodrell Bank Centre for Astrophysics, University of Manchester, \\ Turing Building, Oxford Road, Manchester M13 9PL, UK \\
$^{15}$Centre for Astrophysics, University of Hertfordshire, Hatfield, Herts, UK     \\
$^{16}$Observatoire de Paris, 5 Place Jules Janssen, 92195 Meudon, France   \\
$^{17}$Radboud University Nijmegen, Heijendaalseweg 135,  6525 AJ Nijmegen, The Netherlands \\
$^{18}$Department of Physics and Astronomy, The Open University, Milton Keynes, 
MK7 6AA,\\ Space Science and Technology Department,\\ STFC Rutherford Appleton Laboratory, Chilton, OX11 0QX, UK
}


\pubyear{xxxx}
\volume{xx}
\date{Received xxx; accepted xxx}
\maketitle
\label{firstpage}
\begin{abstract}
At very low frequencies, the new pan-European radio telescope LOFAR is opening the last unexplored window of the electromagnetic spectrum for astrophysical studies. The revolutionary APERTIF phased arrays that are about to be installed on the Westerbork radio telescope (WSRT) will dramatically increase the survey speed for the WSRT. Combined surveys with these 
two facilities will deeply chart the northern sky
over almost two decades in radio frequency from $\sim 15$ up to 1400 MHz.  Here we briefly describe some of the capabilities 
of these new facilities 
and what radio surveys are planned to study fun-damental issues related the formation and evolution of galaxies and clusters of galaxies.
 In the second part we briefly review some recent observational results directly showing that 
diffuse radio emission in clusters traces shocks due to cluster mergers. As these diffuse radio sources are relatively bright at low 
frequencies, LOFAR should be able to detect thousands of such sources up to the epoch of cluster formation. This will 
allow addressing many question about the origin and evolution of shocks and magnetic fields in clusters. At the end we 
briefly review some of the first and very preliminary LOFAR results on clusters. 

\end{abstract} 

\begin{keywords}
Galaxies: clusters: general, intracluster medium; Radio continuum: galaxies; Radio telescopes
\end{keywords}

\section{Introduction}

At low frequencies, the new pan-European radio telescope
LOFAR is opening the last unexplored window of the electromagnetic
spectrum for astrophysical studies. The revolutionary APERTIF phased
arrays that are about to be installed on the Westerbork radio
telescope (WSRT) will dramatically increase the survey speed for the
WSRT. The resulting vast area of new observational parameter space
will be fully exploited for many studies directly related to the
formation of massive black holes, galaxies, and clusters. Particularly
important are three research areas that are driving the design of
several surveys that are planned to be carried out with these new
facilities. These areas are: (i) forming massive galaxies at the epoch of
reionisation, (ii) magnetic fields and shocked hot gas associated with
the first bound clusters of galaxies, and (iii) star formation
processes in distant galaxies. Furthermore, a most exciting
aspect of LOFAR is that its enormous instantaneous field of view coupled
with its unprecedented sensitivity at low frequencies equips LOFAR for
the discovery of new classes of rare extreme-spectrum sources.

In this contribution, we will first briefly describe LOFAR and
APERTIF. For a more extended description of LOFAR we refer to the
contribution of George Heald that extensively describes LOFAR and the
way the data will be handled to form deep images at low frequencies.
Second, we will discuss how  the prime science drivers led to the definition of the planned continuum surveys
with LOFAR and APERTIF. Third, we briefly review some of the work we
have been carrying out to understand diffuse radio emission associated
with merging clusters.  We will mainly concentrate on some of the
statistical results obtained for a  partly new sample of relics showing
correlations between their sizes, spectral indices and distances from
the clusters centres. In the contribution of Reinout van Weeren, he will
high-light recent results for newly discovered individual relics, including the recently discovered 
spectacular double relics in the cluster CIZA J2252.8+5301 (van Weeren et al. 2010). \nocite{wee10a} Finally, a few preliminary results
from LOFAR observations mainly related to clusters are briefly presented. These results show the enormous potential that LOFAR 
has for studying shocks and magnetic fields in clusters.

\section{LOFAR} 

LOFAR, the Low Frequency Radio Array, is a pan-European radio
telescope that is currently being commissioned.  Its revolutionary
design makes use of phased array technology. This replaces the
traditional and expensive mechanical dishes by a combination of simple
receivers and modern computing equipment. LOFAR has two types of
antennas, one optimised for the 30 - 80 MHz range and one for the 110
- 240 MHz range. The antennas are grouped together in stations the
size of soccer fields. The signals from the antennas will be digitised
so that many beams on the sky can be formed. This  makes LOFAR an extremely
efficient instrument to survey large areas of sky. The Dutch part of
the array will be finished in 2011 and will comprise 40
stations distributed over an area of diameter of 100 km. In addition,
in 2011 eight stations in a number of European countries (Germany, UK,
Sweden, and France) are planned to  be operational.  Currently many
functional elements of the LOFAR imaging system are in place. These
elements include: (i) station beam formation, (ii) high speed data transport,
(iii) software correlator to produce visibilities, (iv) calibration algorithms, and (v) wide field map
making. Although a significant amount
of both continued commissioning and technical research will be needed
to obtain maps with the theoretical noise levels, the maps that are currently
produced already are the deepest ever at these low frequencies. 

With its unique design, LOFAR will provide enormous improvements over
previous facilities in the following three regions of parameter space:

\begin{itemize} 

\item Very Low Frequencies, with 2 - 3 orders of magnitude improvement
  in both sensitivity and angular resolution. This is a mostly
  unexplored spectral region that is uniquely sensitive to ultra-steep
  spectrum $z>6$ radio galaxies, diffuse emission from clusters and the
  oldest `fossil' synchrotron electrons.

\item Size of the Instantaneous Field of View, of many tens of square
  degrees. This will deliver a transformational increase in speed to
  survey the radio sky, crucially important for the quest for rare
  objects such as distant clusters, proto-clusters and $z > 6$ radio
  galaxies and rare transient phenomena. 

\item Low-Frequency Radio Spectroscopy, enabling studies of redshifted
  neutral hydrogen at the Epoch of Reionisation.

\end{itemize} 
The design of LOFAR is very versatile and has led to the development of 6 key science projects,
related to cosmic rays, epoch of reionisation, transients and pulsars, cosmic magnetism, the Sun, 
and extragalactic surveys. In this contribution, we will focus on LOFAR and APERTIF surveys to probe 
the extragalactic sky.

\section{APERTIF} 

APERTIF, the new Phased Array Feed receiver system for the Westerbork
Synthesis Radio Telescope (WSRT) will dramatically enlarge the
instantaneous Field-of-View of the WSRT (see Oosterloo et al 2010
\nocite{oos10} for a detailed description). This is done by replacing
the current single Frontend Feeds with Phased Array Feeds (PAFS). Each
of the PAFs consists of 121 Vivaldi elements and will detect the
radiation field (in dual polarisation) in the focal plane of each dish
over an area of about one square meter. Because of this, many beams
can be formed simultaneously for each dish making it possible to image
an area of about 8 square degree on the sky, which is an increase of
about a factor 30 compared to the current WSRT. Its large 300 MHz
bandwidth will not only cater for sensitive continuum imaging, but is also crucial 
for  efficient HI and OH emission surveys and for 
studies of polarised emission from large areas.

\section{Transformational radio surveys} 

\subsection{LOFAR} 

\begin{figure}[h!]
\centerline{
\includegraphics[width=12cm,angle=0]{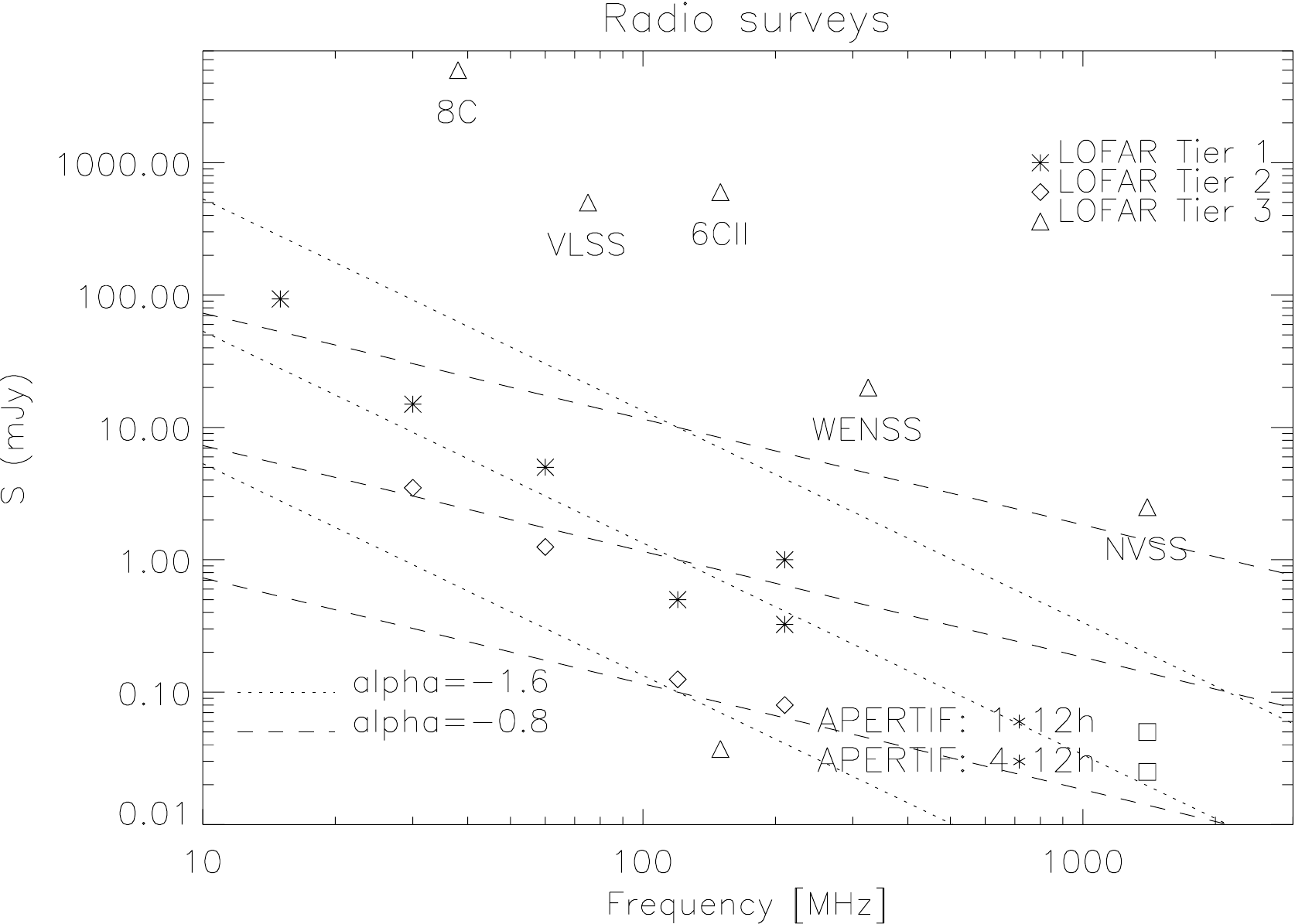}}
\caption{\label{other} \it Flux limits (5 sigma) of the proposed LOFAR and APERTIF 
surveys compared to other existing radio surveys.  The triangle represent
existing surveys: HDF (VLA Richards et al. 2000; WSRT Garrett et
al. 2000), WENSS, NVSS, 6C, VLSS and 8C.  The lines represent different
power-laws ($S\sim \nu^{\alpha}$, with $\alpha=-1.6$ and $-0.8$) to
illustrate how, depending on the spectral indices of the sources, the
LOFAR surveys will compare to other surveys.  }
\end{figure}

The three fundamental areas of astrophysics that have driven the
design of the planned LOFAR surveys are: (i) forming massive galaxies at the epoch of
reionisation, (ii) magnetic fields and shocked hot gas associated with
the first bound clusters of galaxies, and (iii) star formation
processes in distant galaxies.
The areas, depths and
frequencies of the surveys have been chosen so that they would
contain: (i) ~100 powerful radio galaxies close to or at the epoch of
reionisation, (ii) ~100 radio halos at the epoch when the first
massive bound galaxy clusters appeared, and (iii) ~100 proto-clusters.
The resulting  survey parameters are based on estimates of luminosity functions
for powerful radio galaxies by Wilman et al. 2008, \nocite{wil08} for
radio halos by En\ss lin \& R\"ottgering 2002 and Cassano et al. 
2010, and for proto-clusters by Venemans et al. 2007. \nocite{ens02b,cas10a,ven07}  
To achieve the goals of the LOFAR surveys,
a three-tiered approach has been adopted (for details see
R\"ottgering et al. 2010). \nocite{rot10} Tier-1 represents the all sky survey at
frequencies 15, 30, 60 and 120 MHz. Tier-2 are the medium deep surveys
over 1000 sqr deg at 30, 60, 120 and 200 MHz, while Tier-3 encompasses
about 100 sqr degrees down to an extreme depth of  6 $\mu$Jy rms at 150 MHz. The resulting depth
versus frequency is given in Fig. \ref{other}. In addition, very deep data will
be taken on a selected sample of 60 nearby clusters.  Another important 
motivation of LOFAR is to provide the entire international
astronomical community with unique surveys of the radio sky that have
a long-lasting legacy value for a broad range of astrophysical
research. The international LOFAR survey team has identified a range
of fundamental astrophysical research topics on which LOFAR surveys
will have an important impact. These topics include (i) the formation
and evolution of large scale structure of the Universe, (ii) the
physics of the origin, evolution and end-stages of radio sources, (iii)
the magnetic field and interstellar medium in nearby galaxies, and
(iv) Galactic sources such as supernova remnants, HII regions,
exoplanets and pulsars.

\nocite{ric00,gar00}

\subsection{WODAN} 
The extremely large field of view of APERTIF 
would enable the WODAN (Westerbork Observations of the Deep APERTIF Northern-Sky)
project. This project  aims to chart the entire accessible northern sky at 1400
MHz down to 10 $\mu$Jy rms and about 1000 deg$^2$ down to 5  $\mu$Jy. 
WODAN will be an important compliment to the EMU
(Evolutionary Map of the Universe) project.  EMU will use the phased-array feed (PAF) 
mounted on the Australian SKA Pathfinder (ASKAP, De-boer et al.  2009) \nocite{deb09} 
to chart the entire sky south of $\delta = 30^\circ$
to a similar depth as WODAN. For a detailed description of EMU we will refer to 
the contribution of Ray Norris to this conference. 

WODAN and EMU have an enormous synergy with the
LOFAR surveys: virtually all the $5 \times 10^7$ radio sources from
the LOFAR all sky surveys will have their flux density at 1400 MHz
measured. It will yield
radio data for all radio loud AGN, and most luminous starbursts up to $z=2$. 
The resulting densely populated radio color-color diagrams
will be a powerful tool to spectrally discriminate between very rare
radio sources with extreme radio spectra such as diffuse emission from
clusters and very distant radio galaxies. For nearby resolved sources
it will instantly yield spectral index and spectral curvature maps, a
very rich source of information to constrain many physical
parameters. As the combined surveys will cover the entire sky, measurements of  the Integrated
Sachs-Wolfe effect,  galaxy auto-correlation functions and
cosmic magnification will 
significantly tighten cosmological model parameters
(Raccanelli et al. submitted). 

\section{LOFAR and diffuse radio emission from clusters of galaxies} 

Clusters of galaxies are large ensembles of hundreds of galaxies
embedded in hot gas and held together by gravity. 
Besides the hot thermal gas observed in X-rays, the
intra-cluster-medium (ICM) contains relativistic electrons (E $\approx$ Gev)
and magnetic fields ($1 -10 \mu$G), which have been detected via
synchrotron emission in the radio band. LOFAR is uniquely suited to
probe these synchrotron emitting regions and will address many
questions related to the large-scale magnetic fields and relativistic
particles mixed with the thermal ICM. These questions include: What
are the strengths and topologies of the magnetic fields? When and how
were the first magnetic fields generated? How were magnetic fields
subsequently amplified and maintained? 

Furthermore, diffuse radio
sources in galaxy clusters are likely to be direct signatures of huge
shock waves caused by massive cluster mergers. These shocks have a
crucial impact on the energetics and detailed temperature distribution
of the cluster gas. LOFAR observations are therefore very relevant for
studies of the evolution of the energy content of both the thermal and
non-thermal gas in the cluster.  Some of the most prominent nearby
clusters of galaxies host such diffuse synchrotron emitting radio
sources. Classical examples of spectacularly large ($\sim 1$ Mpc) diffuse
cluster emission have been found for the Coma cluster, Abell 2256 and
Abell 3667. The properties of the associated clusters are extreme:
they are very X-ray luminous, have high temperatures (kT $> 7$ keV),
large masses ($> 10^{15}$ M$_\odot$), and high galaxy velocity
dispersions. The overall properties are indicative of the violent
merging of sub-clusters, an important process in the assembly of
massive clusters.  Diffuse radio emission associated with clusters of
galaxies has been classified into three groups: relics, halos and
phoenixes (e.g. Giovannini \& Feretti 2004). \nocite{gio04} 

{\it Cluster relics} are large elongated diffuse structures at the periphery
of clusters. Recently we have discovered a spectacularly long and
narrow relic with a size of 2.0 Mpc $\times$ 50 kpc, located at a
distance of 1 Mpc from the centre of the merging cluster CIZA
J2242.8+5301 (van Weeren et al. 2010). \nocite{wee10a} The relic displays highly aligned magnetic fields and a strong
spectral index gradient due to cooling of the synchrotron emitting
particles in the post shock region. We have argued that these
observations provide conclusive evidence that shocks in merging
clusters produce extremely energetic cosmic rays. Detailed modelling
of the morphology, polarization properties and variations of the radio
spectrum, allowed us to determine the strength of the magnetic field
(5 $\mu$G) and the Mach number ($4.6^{+1.3}_{-0.9}$) of the shock. Our numerical
simulations indicated that the impact parameter of the cluster
collision was about zero and the mass ratio of the colliding clusters
was roughly 2:1 (van Weeren et al. in prep.). 

{\it Cluster radio halos} are located at the centres of clusters, their
diffuse morphologies following that of the X-ray emission. The origin
of the halos is not understood. Especially their enormous $\sim 1$ Mpc sizes
pose problems. The radiative lifetimes of the synchrotron emitting
electrons are so short that the electrons need to have been accelerated
to relativistic speeds close to the place where they radiate. Although
many explanations have been put forward, a currently favoured one is
that turbulence due to cluster mergers is capable of accelerating
electrons to relativistic speeds (e.g. Brunetti et
al. 2001). \nocite{min01} Alternatively, relative electrons could be produced 
when 
relativistic protons from  AGNs in the cluster collide with thermal protons within the cluster gas (Dennison, 1980).  
A second important issue
relates to the origin of the magnetic fields (e.g. Dolag 2006). \nocite{dol06} Are
they primordial in origin and have turbulent processes subsequently
amplified the fields? Or have outflows from active galaxies or
starburst galaxies transported magnetic fields into the inter-galactic
medium? 

{\it Radio phoenixes} are suggested to be due to shocks in the cluster gas
that would adiabatically compress old radio plasma ejected by former
active galaxies. The resulting diffuse objects would have an extremely
steep radio spectrum making them relatively bright at low radio frequencies
(En\ss lin and Kopal-Krishna 2001). \nocite{ens01} Simply considering the timescales
related to the AGN activity, synchrotron losses, and the presence of
shocks we recently argued that such sources could determine the
general appearance of clusters in low frequency LOFAR maps (van Weeren,
et al. 2009a). \nocite{wee09b} 

Because these radio sources associated with cluster wide shocks are diffuse, have low luminosities and steep radio
spectra, they are difficult to detect with conventional radio
observatories, such as Westerbork. As a result there are only about 50
cluster radio sources currently known. Due to its extreme sensitivity
at low radio frequencies, LOFAR will be the break- through instrument
for this field of research (Cassano et al. 2010). \nocite{cas10} For the first time, the
occurrence and characteristics of diffuse cluster radio sources will
be measured as a function of cluster properties up to the epoch at
which the first massive clusters assembled ($z \sim 1$). This will directly
show the effects of shock waves on the evolution of the cluster gas
and magnetic fields, and test predictions that cluster merging is
rampant at high redshift. Detailed LOFAR maps of rotation measures,
polarization properties and radio spectra of nearby halos will
distinguish between the various physical models for the origin of the
diffuse radio emission. It also will probe radio AGN activity over
long time scales, important for studies of the radio feedback
processes in clusters. With the  LOFAR observations, we will
address the following questions:

\begin{itemize} 
\item What are the properties of the cluster-wide shocks (rate of occurrence, volume filling, geometry, Mach numbers)? How do they accelerate particles?
\item What are the characteristics of the magnetic fields (strength, topology)? And how do these relate to models of the origin of the fields?
\item What is the total energy input into the cluster medium by radio loud AGN?
\item What are the properties of the merging clusters (mass ratios, impact parameters) as can be directly deduced by the relic morphologies?
\item How do the properties of merging clusters evolve over cosmic time?
\end{itemize} 

\nocite{cla06}

\section{Towards a sample of relics, a prelude to LOFAR}

As discussed, detailed  radio observations of individual relics clearly suggest that relics originate in  
shocks induced by merging clusters. This scenario can be further tested by studying larger  samples of relics. 
From  GMRT, WSRT and VLA observations of a sample of diffuse radio sources from the 74 MHz VLSS 
survey with spectral indices $\alpha < -1.7$, 5 new relics were discovered. A comparison of the NVSS and WENSS 
radio catalogues with the ROSAT all sky catalogue, 5 additional relics were found. Combined with 17 known relics 
from the literature, the resulting sample was large enough for a statistical study. For details we refer to van Weeren et al. 
2009b. \nocite{wee09c}
For this sample, 
we found that larger relics are mostly located in the cluster periphery, while smaller relics are found closer to the cluster center. 
We also discovered an anti-correlation between the steepness of the spectral index and
the physical size of the relics. A likely explanation for these two 
correlations is that the larger shock waves occur mainly in
lower-density regions. The larger shocks then have larger Mach numbers translating into flatter radio spectra. 
As larger relics are also more luminous, this then also explains that 
within this sample the 
more luminous radio relics have flatter spectral indices. 
Finally, there is a tendency for the steep spectrum relics to show more spectral curvature. 
This would provide evidence for spectral ageing due to inverse compton and/or synchrotron losses. 
We note however that some of the smallest relics might be due to the compression of fossil AGN radio plasma. 
Their very steep and curved spectrum sources are also consistent with this scenario.

\subsection{LOFAR and cluster observations: the rich cluster of galaxies Abell 2256} 

Abell 2256 is a rich X-ray cluster at $z = 0.058$ that has undergone a
merging event estimated to have happened 0.3 Gyr ago (e.g. Miller et
al 2003). \nocite{mil03} Apart from 9 tailed sources, it rather
exceptionally contains three classes of diffuse cluster radio sources:
relics, halos and phoenixes. The northern relics have been discovered
a long time ago (Bridle and Fomalon 1976) \nocite{bri76} and were
studied in detail by Clarke and En\ss lin (2006). \nocite{cla06} They
also clearly showed that A2256 possesses a central halo with a
luminosity following the X-ray - radio halo luminosity relation (Liang
et al. 2000). \nocite{lia00} In very deep 325 MHz GMRT radio maps van
Weeren et al. (2009a) \nocite{wee09b} recently discovered three diffuse
elongated radio sources with extremely steep spectral indices located
about 1~Mpc from the cluster center. These properties indicate that
these objects can be classified as phoenixes.

As A2256 is one of the most luminous radio emitting clusters showing
so many intriguing characteristics, it was one of the prime candidates
to be observed during early commissioning of LOFAR (see also R\"ottgering et al. 2010; Heald et al. 2010). \nocite{hea10} It was observed in the
HBA band in May 2010 for about 8 hours. The data were taken with 10
core stations and 5 remote stations and the observed frequencies
ranged from 115 to 165 MHz.  An image from 18 subbands covering a
total of 4 MHz of bandwidth around 135 MHz was made (see Figure
\ref{a2256}).  The resolution of the image is $31 \times 19$ arcsec
and the noise is $\sim 5$ mJy/beam. So far, the deepest images at low
frequencies have been obtained with the GMRT at 150 MHz (Intema 2009,
Intema et al. submitted, Kale and {Dwarakanath
2010). \nocite{int09b,kal10} The GMRT image clearly shows the relic
and several of the head tail galaxies that are also visible on the
LOFAR image. The GMRT image recovers the central part of the halo
emission.  With LOFAR's very sensitive central core, the full extent
of the halo is visible, showing LOFAR's power to study diffuse steep
spectrum emission from clusters.  Next steps in improving this image
are reducing the data from all the 256 sub-bands, and the
application of more sophisticated data reduction algorithms. These
include proper wide-field imaging taking the varying station beams
into account, iteration of self-calibration/peeling loops, and
removal of ionospheric corrections following the ``SPAM'' method
(Intema et al. 2009). \nocite{int09a} Finally, recently we have observed 
A2256 in the lowband and produced images at  20, 30 and 49 MHz. A spectral map 
from a combination of 49 MHz and the 350 MHz WSRT data (Brentjens et al. 2008)
\nocite{bre08} very nicely spatially resolved the extremely streep spectrum central radio halo 
from the flatter spectrum northern relics. We also recently obtained data on Coma A and A2255. The 
commissioning team is working very hard to obtain excellent images. 

\bigskip 

\noindent 
{\it Acknowledgements} \quad  
C. Ferrari acknowledges financial support by the Agence Nationale
de la Recherche through grant ANR-09-JCJC-0001-01.

\begin{figure} \centerline{
 \includegraphics[width=10cm]{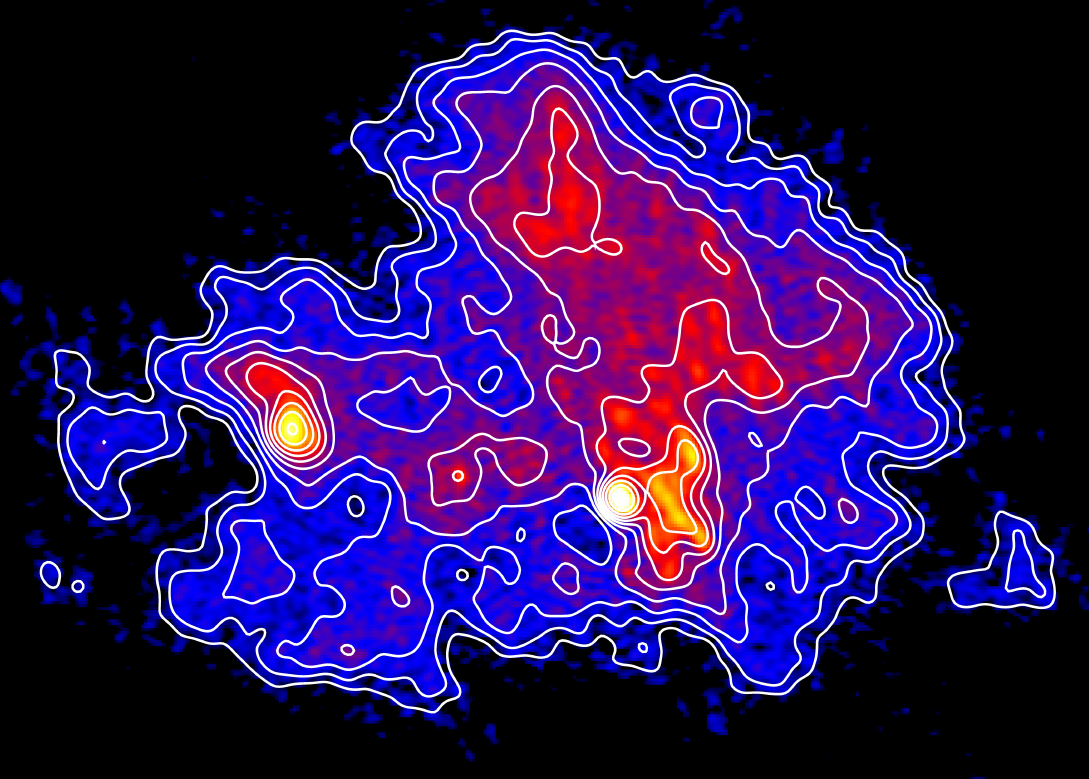}}
\caption{\label{a2256} A 135 MHz LOFAR image of the rich X-ray cluster
  of galaxies A2256 at $z = 0.058$. The image has been made from data
  that were taken with 10 core stations and 5 remote stations. Only 4
  MHz bandwidth out of a total 48 MHz was used during the reduction.
  The resolution of the image is about $31 \times 19$ arcsec and the
  noise is $\sim 5$ mJy/beam. Beside several tailed galaxies and the
  relic structures, the image shows for the first time a spatially
  resolved central halo of A2256 at low frequencies.  }
\end{figure}


\end{document}